\documentclass[12pt]{article}
\usepackage{amssymb,amsmath,epsfig}
\allowdisplaybreaks
\begin{document}
\title{\bf Polarization Modes of Gravitational Wave for Viable $f(R)$ Models}
\author{M. Sharif \thanks{msharif.math@pu.edu.pk} and Aisha Siddiqa
\thanks{aisha.siddiqa17@yahoo.com}\\
Department of Mathematics, University of the Punjab,\\
Quaid-e-Azam Campus, Lahore-54590, Pakistan.}

\date{}

\maketitle
\begin{abstract}
In this paper, we study the gravitational wave polarization modes
for some particular $f(R)$ models using Newman-Penrose formalism. We
find two extra scalar modes of gravitational wave (longitudinal and
transversal modes) in addition to two tensor modes of general
relativity. We conclude that gravitational waves correspond to class
$II_{6}$ under the Lorentz-invariant E(2) classification of plane
null waves for these $f(R)$ models.
\end{abstract}

\section{Introduction}

Gravitational waves (GWs) are fluctuations in the fabric of
spacetime produced by the motion of massive celestial objects. The
scientific curiosity and struggles to detect these waves by the
Earth based detectors lead to the invention of laser interferometer
detectors such as LIGO, VIRGO, GEO and LISA (Bassan 2014). The most
promising source for these detectors is merging the compact binaries
composed of neutron star-neutron star, neutron star-black hole and
black hole-black hole. These orbiting systems loose their energy in
the form of GWs which speed up their orbital motion and this process
ends up at the merging of orbiting objects. Recently, LIGO
scientific and Virgo collaborations (Abbott \textit{et al}. 2016)
detected these waves and provided two observational evidences (with
signals known as GW150914 and GW151226) for GWs each of which is the
result of a pair of colliding black holes.

Polarization of a wave gives information about the geometrical
orientation of oscillations. A common method to discuss polarization
modes (PMs) of GWs is the linearized theory consisting of metric
perturbations around Minkowski background. Newman and Penrose (1962)
introduced tetrad and spinor formalism in general relativity (GR) to
deal with radiation theory. Eardley \textit{et al}. (1973) used this
formalism for linearized gravity and showed that six Newman-Penrose
(NP) parameters for plane null waves represent six polarization
modes (amplitudes) of these GWs. They also introduced
Lorentz-invariant E(2) classification of plane null waves.

Hawking (1971) found an upper bound for the energy of gravitational
radiation emitted by the collision of two black holes. Wagoner
(1984) investigated gravitational radiation emitted by accreting
neutron stars. Culter and Flanagan (1994) explored the extent of
accuracy of the distance to source and masses as well as spin of two
bodies measured by the detectors LIGO and VIRGO from the
gravitational wave signal. Turner (1997) worked on GWs produced by
inflation and discussed the potential of cosmic microwave background
anisotropy as well as laser interferometers (LIGO, VIRGO, GEO and
LISA) for the detection of GWs. Langlois \textit{et al}. (2000)
studied the evolution of GWs for a brane embedded in
five-dimensional anti-de Sitter universe and showed that a discrete
normalizable massless graviton mode exists during slow roll
inflation.

Recent indications of accelerated expansion of the universe caused
by dark energy introduced much interest in cosmology. The mysteries
of dark energy and dark matter (invisible matter) leads to modified
theories of gravity obtained by either modifying matter part or
geometric part of the Einstein-Hilbert action. A direct
generalization of GR is the $f(R)$ theory in which the Ricci scalar
$R$ in the Einstein-Hilbert action is replaced by its generic
function $f(R)$. De Felice and Tsujikawa (2010) presented a
comprehensive study on various applications of f(R) theory to
cosmology and astrophysics. Starobinsky (1980) proposed the first
inflationary model in $f(R)$ gravity compatible with anisotropies of
cosmic microwave background radiation. Hu and Sawicki (2007)
proposed a class of $f(R)$ models without cosmological constant that
satisfy cosmological and solar system tests for small field limit of
the parameter space. Tsujikawa (2008) explored observational
consequences of $f(R)$ models that satisfy the local gravity
constraints. Bamba {\it et al.} (2010) introduced $f(R)$ model which
explains inflation and late cosmic expansion at the same time.

A lot of work has been done for PMs of GWs in $f(R)$ as well as in
other modified theories. Capozziello \textit{et al}. (2008)
investigated PMs of GWs in $f(R)$ gravity and concluded that for
every $f(R)$ model there is an extra mode than GR called massive
longitudinal mode. They also worked out the response function of GWs
with LISA. Alves \textit{et al}. (2009) discussed PMs of GWs for
particular $f(R)$ model concluding the same results. They showed
that five non-zero PMs exist for a specific form of quadratic
gravity. The topic of gravitational radiation for linearized $f(R)$
theory has also been discussed in literature (Berry and Gair 2011;
N$\ddot{a}$f and Jetzer 2011). Capozziello and Stabile (2015)
studied GWs in the context of general fourth order gravity and
discussed the states of polarization and helicity. Kausar \textit{et
al}. (2016) found that for any $f(R)$ model there are two extra
modes as compared to GR. Alves \textit{et al}. (2016) explored these
modes in $f(R,T)$ as well as $f(R,T^{\phi})$ theories concluding
that the earlier one reduces to $f(R)$ in vacuum while PMs for the
later depend on the expression of $T^{\phi}$.

Herrera {\it et al.} (2015a, 2015b) studied the presence of
gravitational radiation in GR for perfect as well as dissipative
dust fluid with axial symmetry using super-Poynting vector and
showed that both fluids do not produce gravitational radiation. We
have investigated that the axial dissipative dust acts as a source
of gravitational radiation in $f(R)$ theory (Sharif and Siddiqa
2017). This paper is devoted to find PMs for some viable dark energy
models of this gravity. The paper is organized as follows. In the
next section, we write down field equations and dark energy models
of $f(R)$ gravity. We then find PMs of GWs for three models in its
subsections. Finally, we conclude our results.

\section{Dark Energy Models in $f(R)$ Gravity}

The $f(R)$ gravity action is defined as
\begin{equation}\label{1}
S=\frac{1}{16\pi}\int\sqrt{-g}f(R)d^{4}x+S_{M},
\end{equation}
where $S_{M}=\int\sqrt{-g}L_{M}d^{4}x$ denotes the matter action and
$L_{M}$ represents the matter Lagrangian. To discuss PMs of GWs, one
needs to investigate the linearized far field vacuum field
equations. The vacuum field equations for the action (\ref{1}) are
given by
\begin{equation}\label{2}
F(R)R_{\beta\gamma}-\frac{1}{2}f(R)g_{\beta\gamma}-
\nabla_{\beta}\nabla_{\gamma}F(R)+g_{\beta\gamma}\Box F(R)=0,
\end{equation}
where $F=\frac{df}{dR}=f_{R}$ and
$\Box=\nabla^{\alpha}\nabla_{\alpha}$ is the D'Alembertian operator.
The trace of this equation is
\begin{equation}\label{3}
RF(R)-2f(R)+3\Box F(R)=0.
\end{equation}
We assume that waves are traveling in $z$-direction, i.e., each
quantity can be a function of $z$ and $t$.

Various models of $f(R)$ gravity have been proposed in literature
(Starobinsky 1980; Hu and Sawicki 2007; Tsujikawa 2008; Bamba
\textit{et al}. 2010) describing the phenomena of early inflation
and late cosmic expansion. The model proposed by Hu and Sawicki
(2007) is reduced to the model considered by Alves \textit{et al.}
(2009) in the weak field regime (i.e., when $R<<m^{2}$, $m$ stands
for mass). Thus Hu and Sawicki model which satisfies the
cosmological and solar system tests has been indeed examined for PMs
of GWs in the low curvature case or Minkowski background. Similarly,
the Starobinsky model having consistency with the temperature
anisotropies measured by CMBR (De Felice and Tsujikawa 2010) has
also been analyzed for PMs of GWs by Kausar \textit{et al}. (2016).

Amendola \textit{et al}. (2007) derived the conditions for
cosmological viability of some dark energy models in $f(R)$ gravity.
They divided $f(R)$ models into four classes according to the
existence of a matter dominated era and the final accelerated
expansion phase or geometrical properties of the $m(r)$ curves where
$m(r)=\frac{Rf_{,RR}}{f_{,R}}$. They concluded that models of class
I are not physical, class II models asymptotically approach to de
Sitter universe, class III contains models showing strongly phantom
era and models of class IV represent non-phantom acceleration
$(\omega>-1)$. They argued that only models belonging to class II
are observationally acceptable with the final outcome of
$\Lambda$CDM model. Here we consider these observationally
acceptable models having the similar geometry of $m(r)$ curves to
discuss the PMs of GW. There are four models among the considered
models that fall in class II while the model $R+\alpha R^{-n}$ has
already been discussed by (Alves \textit{et al}. 2009) so we discuss
the remaining three models in this paper.

\subsection{Polarization Modes for $f(R)=R+\xi R^{2}-\Lambda$}

We consider the model $f(R)=R+\xi R^{2}-\Lambda$, it is assumed that
$\xi$ (an arbitrary constant) and $\Lambda$ (cosmological constant)
have positive values (Amendola \textit{et al}. 2007). This model
corresponds to $\Lambda$CDM model in the limit $\xi\rightarrow0$ and
Starobinsky inflationary model for $\Lambda\rightarrow0$. In this
case, Eq.(\ref{3}) yields
\begin{equation}\label{4}
R(1+2\xi R)-2(R+\xi R^{2}-\Lambda)+3\Box (1+2\xi R)=0,
\end{equation}
which on simplification gives
\begin{equation}\label{5}
\Box R-\frac{1}{6\xi}R=-\frac{\Lambda}{3\xi}.
\end{equation}
For the sake of simplicity, we consider gravitational waves moving
in one direction, i.e., $z$-direction. Thus Eq.(5) can be
interpreted as a non-homogeneous two-dimensional wave equation or
Klein-Gordon equation and its solution can be found using different
methods like Fourier transform and Green's function etc. Here we
obtain its solution following the technique used to solve
Klein-Gordon and Sine-Gordon equations given in (Rajaraman 1986)
which is simple as compared to other methods. According to this
method, any static solution is a wave with zero velocity and for the
systems with Lorentz invariance, once a static solution is known,
moving solutions are trivially obtained by boosting, i.e.,
transforming to a moving coordinate frame. Since we are considering
the vacuum field equations and the background metric is Minkowski,
so we can apply Lorentz transformations to the Ricci scalar $R$ (a
Lorentz invariant quantity). Hence static solution of Eq.(\ref{5})
is obtained by solving
\begin{equation}\label{6}
\frac{d^{2}R}{dz^{2}}-\frac{1}{6\xi}R=-\frac{\Lambda}{3\xi},
\end{equation}
whose solution is
\begin{equation}\label{7}
R(z)=c_{1}e^{\sqrt{m}z}+c_{2}e^{-\sqrt{m}z}+2\Lambda,
\end{equation}
where $c_{1}$, $c_{2}$ are constants of integration and
$m=\frac{1}{6\xi}$.

Since our system is Lorentz invariant, the time dependent solution
is obtained from the static solution through Lorentz transformation
as
\begin{equation}\label{8}
R(z,t)=c_{1}e^{\sqrt{m}\frac{z-v
t}{\sqrt{1-v^{2}}}}+c_{2}e^{-\sqrt{m}\frac{z-v
t}{\sqrt{1-v^{2}}}}+2\Lambda,
\end{equation}
where $\sqrt{1-v^{2}}$ is the Lorentz factor and $v$ represents the
velocity of wave propagation. Also, Eq.(\ref{2}) can be rewritten as
\begin{equation}\label{9}
R_{\beta\gamma}=\frac{1}{F(R)}\left[\frac{1}{2}f(R)g_{\beta\gamma}+
\nabla_{\beta}\nabla_{\gamma}F(R)-g_{\beta\gamma}\Box F(R)\right].
\end{equation}
Replacing the values of $f(R)$ and $F(R)$, we obtain its linearized
form as
\begin{equation}\label{10}
R_{\beta\gamma}=\frac{1}{2}(R-\Lambda+2\xi\Lambda
R)g_{\beta\gamma}+2\xi \nabla_{\beta}\nabla_{\gamma}R-2\xi
g_{\beta\gamma}\Box R.
\end{equation}
The non-zero components of Ricci tensor are
\begin{eqnarray}\label{10a}
R_{tt}&=&\frac{1}{6(1-v^{2})}[3v^{2}(R-\Lambda)-(R+3\Lambda)]-\xi\Lambda
R,\\\label{10b} R_{xx}&=&\frac{1}{6}(R+\Lambda)+\xi\Lambda
R=R_{yy},\\\label{10c}
R_{zz}&=&\frac{1}{6(1-v^{2})}[3(R-\Lambda)-v^{2}(R+\Lambda)]+\xi\Lambda
R,\\\label{10d} R_{tz}&=&-\frac{v R}{3(1-v^{2})}.
\end{eqnarray}
With the help of Eqs.(\ref{A.5}) and (\ref{A.9}), we have
\begin{eqnarray}\label{10e}
\Psi_{2}=\frac{1}{12}R,\quad
\Psi_{3}=\frac{1}{2}R_{l\tilde{m}},\quad
\Phi_{22}=-\frac{1}{2}R_{ll}.
\end{eqnarray}

Now, we find the expressions of $\Psi_{3}$ and $\Phi_{22}$ using
Eq.(\ref{A.4}). For $\Psi_{3}$, it yields
\begin{eqnarray}\label{10f}
\Psi_{3}&=&\frac{1}{2}R_{l\tilde{m}}=\frac{1}{2}R_{\mu\nu}l^{\mu}\tilde{m}^{\nu},
\end{eqnarray}
which can also be written as
\begin{eqnarray}\label{10f1}
\Psi_{3}&=&\frac{1}{2}(R_{tt}l^{t}\tilde{m}^{t}+
R_{xx}l^{x}\tilde{m}^{x}+R_{yy}l^{y}\tilde{m}^{y}+
R_{zz}l^{z}\tilde{m}^{z}+R_{tz}l^{t}\tilde{m}^{z}).
\end{eqnarray}
From Eqs.(\ref{A.1}) and (\ref{A.2}), the component form of vectors
$k$, $l$, $m$ and $\tilde{m}$ can be written as
\begin{eqnarray}\label{10f2}
k^{\mu}&=&\frac{1}{\sqrt{2}}(1,0,0,1),\quad
l^{\mu}=\frac{1}{\sqrt{2}}(1,0,0,-1),\\\label{10f3}
m^{\mu}&=&\frac{1}{\sqrt{2}}(0,1,i,0),\quad
\tilde{m}^{\mu}=\frac{1}{\sqrt{2}}(0,1,-i,0).
\end{eqnarray}
Substituting all the required values in Eq.(\ref{10f1}), we obtain
\begin{eqnarray}\label{10f4}
\Psi_{3}=0.
\end{eqnarray}
Similarly, Eq.(\ref{A.4}) for $\Phi_{22}$ yields
\begin{eqnarray}\nonumber
\Phi_{22}&=&-\frac{1}{2}R_{ll}=-\frac{1}{2}R_{\mu\nu}l^{\mu}l^{\nu}=
-\frac{1}{2}(R_{tt}l^{t}l^{t}+2R_{tz}l^{t}l^{z}+R_{zz}l^{z}l^{z}).
\end{eqnarray}
Replacing the Ricci tensor components and components of $l^{\mu}$,
the above equation leads to
\begin{eqnarray}\label{10g}
\Phi_{22}=-\frac{R}{12}\left(\frac{1+v}{1-v}\right)+\frac{\Lambda(2v^{2}+3)}{12(1-v^{2})}.
\end{eqnarray}
Notice that $\Psi_{4}\neq0$ represents the tensor modes of GWs.
Since there is no expression of $\Psi_{4}$ in terms of Ricci tensor,
so it cannot be evaluated with the help of available values of Ricci
tensor and Ricci scalar (Alves \textit{et al}. 2016). It can be
observed that for $\Lambda$CDM model (when $\xi\rightarrow0$)
$\Psi_{2}$ and $\Phi_{22}$ remain non-zero.

The model, $f(R)=R+\xi R^{2}-\Lambda$, is always viable and reduces
to GR when both $\xi$ as well as $\Lambda$ approach to zero. In GR,
there are only two tensor modes of polarization associated with
$Re\Psi_{4}$ and $Im\Psi_{4}$, i.e., we have only $\Psi_{4}$
non-zero among six NP parameters. From Eq.(4), we have $R=0$ for
$\xi\rightarrow0$, $\Lambda\rightarrow0$, hence GR results are
retrieved.

\subsection{Polarization Modes for $f(R)=R^{p}(\ln \alpha R)^{q}$}

This model is observationally acceptable for $p=1$ and $q>0$
(Amendola \textit{et al}. 2007). Here we assume that $q=1$ such that
the model becomes $f(R)=R\ln \alpha R$. Substituting the values of
$f(R)$ and $F(R)$ in Eq.(\ref{3}), it gives
\begin{eqnarray}\label{15}
3\Box \ln \alpha R-R\ln\alpha R+R=0.
\end{eqnarray}
Assuming $\ln \alpha R=\phi$, this equation transforms to
\begin{eqnarray}\label{16}
\Box \phi&=&\frac{e^{\phi}}{3\alpha}(\phi-1),
\end{eqnarray}
which can also be written as
\begin{eqnarray}\label{16a}
\Box \phi&=&\frac{\partial U}{\partial \phi};\quad
U(\phi)=\frac{e^{\phi}}{3\alpha}(\phi-2).
\end{eqnarray}
First we seek for a static solution, i.e., consider $\phi=\phi(z)$
such that integration of Eq.(\ref{16a}) gives
\begin{eqnarray}\label{16b}
\frac{1}{2}\left(\frac{d\phi}{dz}\right)^{2}=U(\phi).
\end{eqnarray}
Substituting the value of $U(\phi)$ and then integrating, it follows
that
\begin{eqnarray}\label{16c}
\phi(z)=2\left[1+\text{InverseErf}\left[\frac{e z}{\sqrt{3\alpha
\pi}}+\frac{ec_{3}}{\sqrt{2\alpha \pi}}\right]^{2}\right],
\end{eqnarray}
where $c_{3}$ is a constant of integration, $e= 2.71828$ and Erf is
defined by
\begin{eqnarray}\label{16c}
Erf(z)=\frac{2}{\sqrt{\pi}}\int_{0}^{z}e^{-s^{2}}ds.
\end{eqnarray}
Using Lorentz transformation, we obtain time dependent solution
given by
\begin{eqnarray}\label{16d}
\phi(z,t)=2\left[1+\text{InverseErf}\left[
\frac{e(z-vt)}{\sqrt{1-v^{2}}\sqrt{3\alpha
\pi}}+\frac{ec_{3}}{\sqrt{2\alpha \pi}}\right]^{2}\right].
\end{eqnarray}
The expression for Ricci scalar is obtained as
\begin{eqnarray}\label{16e}
R(z,t)=\frac{1}{\alpha}\exp\left(2\left[1+\text{InverseErf}\left[
\frac{e(z-vt)}{\sqrt{1-v^{2}}\sqrt{3\alpha
\pi}}+\frac{ec_{3}}{\sqrt{2\alpha \pi}}\right]^{2}\right]\right).
\end{eqnarray}
The non-zero components of the Ricci tensor have the form
\begin{eqnarray}\nonumber
R_{tt}&=&-\frac{R}{6(v^{2}-1)}\left[\frac{(3v^{2}-1)\ln\alpha R
-2}{\ln\alpha R+1}\right],\\\nonumber
R_{xx}&=&\frac{R}{6}\left(\frac{\ln\alpha R+2}{\ln\alpha
R+1}\right)=R_{yy},\\\nonumber R_{tz}&=&\frac{Rv(\ln\alpha
R-1)}{3(v^{2}-1)(\ln\alpha R+1)},\\\nonumber
R_{zz}&=&\frac{R}{6(v^{2}-1)}\left[\frac{(v^{2}-3)\ln\alpha
R+2v^{2}}{\ln\alpha R+1}\right].
\end{eqnarray}
Finally, the NP parameters for this case are
\begin{eqnarray}\label{b1}
\Psi_{2}=\frac{1}{12}R,\quad \Psi_{3}=0,\quad
\Phi_{22}=-\frac{R}{12}\left(\frac{1+v}{1-v}\right)\frac{(\ln\alpha
R-1)}{\ln\alpha R+1}.
\end{eqnarray}
Here $\Psi_{4}$ is also a non-vanishing NP parameter as discussed in
the previous case.

\subsection{Polarization Modes for $f(R)=R^{p}e^{\frac{q}{R}}$}

This model is observationally acceptable for $p=1$, so we take
$f(R)=Re^{\frac{q}{R}}$ (Amendola \textit{et al}. 2007). This model
reduces to GR when $q=0$ and consequently gives no additional PMs.
Thus to find extra PMs, we consider $q\neq0$ in further
calculations. For this model, the trace equation (\ref{3}) becomes
\begin{eqnarray}\label{3a}
Re^{\frac{q}{R}}\left(1-\frac{q}{R}\right)-2Re^{\frac{q}{R}}+
3\Box\left(e^{\frac{q}{R}}\left(1-\frac{q}{R}\right)\right)=0.
\end{eqnarray}
In low curvature regime, we have $R<<q$ which reduces the above
equation to the following
\begin{eqnarray}\label{3b}
\Box\left(\frac{1}{R}e^{\frac{q}{R}}\right)+\frac{1}{3}e^{\frac{q}{R}}=0.
\end{eqnarray}
Replacing $\frac{1}{R}=u$ and $u=u(z)$ for static solution, we
obtain
\begin{eqnarray}\label{3c}
\frac{d^{2}}{dz^{2}}(ue^{qu})+\frac{1}{3}e^{qu}=0.
\end{eqnarray}
Solving the double derivative of the above equation, it becomes
\begin{eqnarray}\nonumber
(1+qu)\frac{d^{2}u}{dz^{2}}+q(qu+2)\left(\frac{du}{dz}\right)^{2}+\frac{1}{3}=0.
\end{eqnarray}
The is a non-homogeneous non-linear second order differential
equation and does not provide an exact analytic solution unless we
make some assumptions to simplify it. Since we are working in the
weak-field regime, so $R$ is very small. Assuming $q$ to be very
large, we have $qu=\frac{q}{R}$ (as $u=\frac{1}{R}$) to be very
large such that $(qu+1)\approx qu$ as well as $(qu+2)\approx qu$ and
the above equation reduces to
\begin{eqnarray}\nonumber
u\frac{d^{2}u}{dz^{2}}+qu\left(\frac{du}{dz}\right)^{2}+\frac{1}{3q}=0.
\end{eqnarray}
Here $\frac{1}{3q}\rightarrow 0$ as $q$ is very large, hence it
reduces to
\begin{eqnarray}\label{3d}
u\frac{d^{2}u}{dz^{2}}+qu\left(\frac{du}{dz}\right)^{2}=0
\end{eqnarray}
whose solution yields $(u=1/R)$
\begin{eqnarray}\label{3f}
R(z)&=&\left(\frac{1}{q}\ln[q(c_{4}z+c_5)]\right)^{-1},
\end{eqnarray}
where $c_{4}$ and $c_{5}$ are integration constants. The non-static
solution becomes
\begin{eqnarray}\label{3g}
R(z,t)&=&\left(\frac{1}{q}\ln\left[q\left(c_{4}\frac{(z-vt)}{\sqrt{1-v^{2}}}
+c_{5}\right)\right]\right)^{-1}.
\end{eqnarray}
The non-vanishing components of the Ricci tensor are
\begin{eqnarray}\nonumber
R_{tt}&=&\frac{q\left[q(z-vt)^{2}+2\ln\left[q\left(c_{4}\frac{(z-vt)}{\sqrt{1-v^{2}}}
+c_{5}\right)\right]\right]c_{4}^{2}}
{2(\ln\left[q\left(c_{4}\frac{(z-vt)}{\sqrt{1-v^{2}}}
+c_{5}\right)\right])^{2}((tv-z)c_{4}-\sqrt{1-v^{2}}c_{5})^{2}}\\\label{29a}
&+&\frac{q\left[2q\sqrt{1-v^{2}}(z-vt)c_{4}c_{5}-qc_{5}^{2}(v^{2}-1)\right]}
{2(\ln\left[q\left(c_{4}\frac{(z-vt)}{\sqrt{1-v^{2}}}
+c_{5}\right)\right])^{2}((tv-z)c_{4}-\sqrt{1-v^{2}}c_{5})^{2}},
\\\nonumber
R_{xx}=R_{yy}&=&\frac{-q\left[q(z-vt)^{2}-2(v^{2}-1)\ln\left[q\left(c_{4}\frac{(z-vt)}{\sqrt{1-v^{2}}}
+c_{5}\right)\right]\right]c_{4}^{2}}
{2(\ln\left[q\left(c_{4}\frac{(z-vt)}{\sqrt{1-v^{2}}}
+c_{5}\right)\right])^{2}((tv-z)c_{4}-\sqrt{1-v^{2}}c_{5})^{2}}\\\label{29b}
&-&\frac{q\left[2q\sqrt{1-v^{2}}(z-vt)c_{4}c_{5}-qc_{5}^{2}(v^{2}-1)\right]}
{2(\ln\left[q\left(c_{4}\frac{(z-vt)}{\sqrt{1-v^{2}}}
+c_{5}\right)\right])^{2}((tv-z)c_{4}-\sqrt{1-v^{2}}c_{5})^{2}},
\\\nonumber
R_{zz}&=&\frac{-q\left[q(z-vt)^{2}-2v^{2}\ln\left[q\left(c_{4}\frac{(z-vt)}{\sqrt{1-v^{2}}}
+c_{5}\right)\right]\right]c_{4}^{2}}
{2(\ln\left[q\left(c_{4}\frac{(z-vt)}{\sqrt{1-v^{2}}}
+c_{5}\right)\right])^{2}((tv-z)c_{4}-\sqrt{1-v^{2}}c_{5})^{2}}\\\label{29c}
&-&\frac{q\left[2q\sqrt{1-v^{2}}(z-vt)c_{4}c_{5}-qc_{5}^{2}(v^{2}-1)\right]}
{2(\ln\left[q\left(c_{4}\frac{(z-vt)}{\sqrt{1-v^{2}}}
+c_{5}\right)\right])^{2}((tv-z)c_{4}-\sqrt{1-v^{2}}c_{5})^{2}},\\\label{29d}
R_{tz}&=&-\frac{qvc_{4}^{2}}{(1-v^{2})\left(\frac{(z-vt)}{\sqrt{1-v^{2}}}c_{4}+c_{5}\right)}.
\end{eqnarray}
The corresponding NP parameters are
\begin{eqnarray}\label{b3}
\Psi_{2}&=&\frac{1}{12}R,\quad \Psi_{3}=0,\\\label{b4}
\Phi_{22}&=&-\frac{Rc_{4}^{2}(v-1)^{2}}{4((tv-z)c_{4}-\sqrt{1-v^{2}}c_{5})^{2}}.
\end{eqnarray}
$\Psi_{4}$ is also non-zero.

\section{Final Remarks}

Observations suggest that our universe is facing an accelerated
expansion phase due to a mysterious factor of dark energy. Moreover,
direct observation of GWs opens up a new window of research. It
would be worthwhile to discuss combine effect of both dark energy
and GWs. In this paper, we have found PMs of GWs in the context of
$f(R)$ dark energy models. For each of the three models, we have
first obtained a static solution of differential equation in $R$ and
then applied Lorentz transformation to obtain a time dependent
solution. It is observed that due to Lorentz transformation, a
factor of $\left(\frac{1+v}{1-v}\right)$ appears in the value of
$\Phi_{22}$ mode in first and second case but it has negligible
effect because the speed of GW is comparable with the speed of
light. It can be seen that the expressions of $\Psi_{2}$ and
$\Phi_{22}$ in all cases are directly proportional to $R$ implying
that increase in $R$ enhances these modes or amplitudes. The mode
$\Phi_{22}$ for the first model (\ref{10g}) depends directly on the
model parameter $\Lambda$, for the second model (\ref{b1}), it
depends on $\ln\alpha$ (as $\ln\alpha R=\ln\alpha+\ln R$) while for
the third model (\ref{b3}), this depends on constants $c_{4}$ and
$c_{5}$.

In each case, we have found four non-zero PMs of GWs $\Psi_{2}$
(longitudinal scalar mode), $\Psi_{4}$ ($+$,$\times$ tensorial
modes) and $\Phi_{22}$ (breathing scalar mode) which is in agreement
with the results of (Kausar \textit{et al}. 2016). We have
non-vanishing $\Psi_{2}$ for each model implying that GWs for $f(R)$
dark energy models correspond to class $II_{6}$ (as mentioned in
Table \textbf{1}). This is the only observer dependent mode,
remaining modes are all observer independent (Eardley \textit{et
al}. 1973). These expressions of NP parameters representing the
amplitudes of GWs are significant due to the presence of dark energy
dominated era.

Here we elaborate the PMs of GW for some modified theories. The six
non-zero PMs are found only for the quadratic gravity with
Lagrangian density $\mathcal{L}=R+\alpha R^{2}+\gamma
R_{\mu\nu}R^{\mu\nu}$ in (Alves \textit{et al}. 2009). For $F(T)$
theory (where T is torsion scalar in teleparallelism), there are no
extra PMs from GR as shown in (Bamba \textit{et al}. 2013) and in
$f(R,T^{\phi})$ theory the number of PMs of GW depend on the
functional form of $f(R,T^{\phi})$ (Alves \textit{et al}. 2016). On
the other hand, the PMs of GW in scalar-tensor theory (Kausar 2017)
and massive Brans-Dicke theory (Sathyaprakash and Schutz 2009) are
same as in $f(R)$ theory.

The LIGO instruments in Livingston and Hanford have similar
orientations and the possibility of extra PMs than GR cannot be
excluded. Moreover, with two detectors having no electromagnetic and
neutrino counterpart, a large uncertainty is expected about the
source of event and consequently in the speed of GW. Thus the
possibility that speed of GW is less than the speed of light cannot
be excluded. Hence from this perspective, the modified theories of
gravity cannot be ruled out. The improvements to automated pipelines
and analysis techniques for the detection of future GW events are
continuously made for accurate measurements. Recently, two more
events of GWs, GW170104 (Abbott \textit{et al}. 2017a) and GW170817
(Abbott \textit{et al}. 2017b) have been detected by the advanced
interferometers. The event GW170104 is consistent with merging black
holes of masses $31M_{\odot}$ and $19M_{\odot}$ in GR while the
second one GW170817 is consistent with the binary neutron star
inspiral having masses in the range $1.17M_{\odot}-1.60M_{\odot}$.
The signal GW170817, has the association with GRB170817A detected by
Fermi-GBM and provides the first direct evidence of a link between
these mergers and short $\gamma$-ray bursts. It is expected that
future GW observations made by a network of the Earth based
interferometers could actually measure the polarization of GWs and
thus constrain $f(R)$ deviations from GR.

\vspace{1.0cm}

{\bf Acknowledgments}

\vspace{0.25cm}

We would like to thank the Higher Education Commission, Islamabad,
Pakistan for its financial support through the {\it Indigenous Ph.D.
5000 Fellowship Program Phase-II, Batch-III}. We are also grateful
to the anonymous referee for his constructive comments.

\renewcommand{\theequation}{A\arabic{equation}}
\setcounter{equation}{0}
\section*{Appendix A}

In this appendix, we first briefly describe the Newman-Penrose
formalism (Newman and Penrose 1962) to discuss gravitational waves
and then PMs as well as classification of null waves is developed
(Eardley \textit{et al}. 1973).

Newman and Penrose developed a new technique in GR with the help of
tetrad formalism and applied this to resolve the issue of outgoing
gravitational radiation. They defined the following relations
between the Cartesian $(\hat{t},\hat{x},\hat{y},\hat{z})$ and
null-tetrads $(k,l,m,\tilde{m})$
\begin{eqnarray}\label{A.1}
k&=&\frac{1}{\sqrt{2}}(\hat{t}+\hat{z}),\quad
l=\frac{1}{\sqrt{2}}(\hat{t}-\hat{z}),\\\label{A.2}
m&=&\frac{1}{\sqrt{2}}(\hat{x}+i\hat{y}),\quad
\tilde{m}=\frac{1}{\sqrt{2}}(\hat{x}-i\hat{y}),
\end{eqnarray}
which satisfy the relations
\begin{eqnarray}\label{A.3}
-k.l=m.\tilde{m}=1,\quad k.m=k.\tilde{m}=l.m=l.\tilde{m}=0.
\end{eqnarray}
Any tensor can be transformed from Cartesian to null basis by the
formula (Alves \textit{et al}. 2009)
\begin{eqnarray}\label{A.4}
S_{abc...}=S_{\alpha\beta\gamma...}a^{\alpha}b^{\beta}c^{\gamma...},
\end{eqnarray}
where $(a,b,c,...)$ vary over the set
$\left\{k,l,m,\tilde{m}\right\}$ and $(\alpha,\beta,\gamma,...)$
vary over the set $\left\{t,x,y,z\right\}$. In (Newman and Penrose
1962), the irreducible parts of the Riemann tensor, also called the
NP parameters, are defined by ten $\Psi$'s, nine $\Phi$'s and a term
$\Lambda$ (these are all algebraically independent). Eardley et al.
(1973) showed that for plane null waves (due to differential and
symmetry properties of the Riemann tensor) these NP quantities are
reduced to the set
$\left\{\Psi_{2},\Psi_{3},\Psi_{4},\Phi_{22}\right\}$. This set
consists of six NP parameters or PMs because $\Psi_{3}$ and
$\Psi_{4}$ are complex and thus represent two independent modes.
They also give formulas of these NP quantities in terms of
null-tetrad components of the Riemann tensor as
\begin{eqnarray}\label{A.5}
\Psi_{2}=-\frac{1}{6}R_{lklk},\quad
\Psi_{3}=-\frac{1}{2}R_{lkl\tilde{m}},\quad
\Psi_{4}=-R_{l\tilde{m}l\tilde{m}},\quad
\Phi_{22}=-R_{lml\tilde{m}}.
\end{eqnarray}
Following are some helpful relations of null-tetrad components of
the Riemann and Ricci tensors
\begin{eqnarray}\label{A.9}
R_{lk}=R_{lklk},\quad R_{ll}=2R_{lml\tilde{m}},\quad
R_{lm}=R_{lklm},~R_{l\tilde{m}}=R_{lkl\tilde{m}},~ R=-2R_{lk}.
\end{eqnarray}
The classification of weak plane null waves (Eardley \textit{et al}.
1973) obtained for standard observer (i.e., each observer sees the
waves traveling in $z$-direction and each observer measures the same
frequency) is given in Table \textbf{1}.\\\\\\\\

Table \textbf{1}: The E(2) Classes of Weak Plane Null Waves
\begin{table}[bht]
\centering
\begin{small}
\begin{tabular}{|c|c|}
\hline Classes&Condition for NP Parameters\\
\hline $II_{6}$&$\Psi_{2}\neq0$\\
\hline $III_{5}$&$\Psi_{2}=0$ and $\Psi_{3}\neq0$\\
\hline $N_{3}$&$\Psi_{2}=0=\Psi_{3}$, $\Psi_{4}\neq0$ and $\Phi_{22}\neq0$ \\
\hline $N_{2}$&$\Psi_{2}=0=\Psi_{3}=\Phi_{22}$ and $\Psi_{4}\neq0$\\
\hline $O_{1}$&$\Psi_{2}=0=\Psi_{3}=\Psi_{4}$ and $\Phi_{22}\neq0$\\
\hline $O_{0}$&$\Psi_{2}=0=\Psi_{3}=\Phi_{22}=\Psi_{4}$\\
\hline
\end{tabular}
\end{small}
\end{table}

\end{document}